\documentclass[epj]{svjour}
\usepackage{graphics}
\usepackage{cite}

\begin{document}

\title{Universality aspects of the 2d random-bond Ising and 3d Blume-Capel models}

\author{N.G. Fytas\inst{1}\thanks{e-mail: nfytas@phys.uoa.gr} \and P.E.
Theodorakis\inst{2,3}
}                     

\institute{Departamento de F\'{i}sica Te\'{o}rica I, Universidad
Complutense, E-28040 Madrid, Spain \and Faculty of Physics,
University of Vienna, Boltzmanngasse 5, 1090 Vienna, Austria \and
Institute for Theoretical Physics and Center for Computational
Materials Science, Vienna University of Technology,
Hauptstra{\ss}e 8-10, 1040 Vienna, Austria}

\date{Received: date / Revised version: date}

\abstract{We report on large-scale Wang-Landau Monte Carlo
simulations of the critical behavior of two spin models in two-
(2d) and three-dimensions (3d), namely the 2d random-bond Ising
model and the pure 3d Blume-Capel model at zero crystal-field
coupling. The numerical data we obtain and the relevant
finite-size scaling analysis provide clear answers regarding the
universality aspects of both models. In particular, for the
random-bond case of the 2d Ising model the theoretically predicted
strong universality's hypothesis is verified, whereas for the
second-order regime of the Blume-Capel model, the expected $d=3$
Ising universality is verified. Our study is facilitated by the
combined use of the Wang-Landau algorithm and the critical energy
subspace scheme, indicating that the proposed scheme is able to
provide accurate results on the critical behavior of complex spin
systems.
}

\authorrunning{N.G. Fytas and P.E. Theodorakis} \titlerunning{Universality aspects of the 2d random-bond Ising and 3d Blume-Capel models}

\maketitle

\section{Introduction}
\label{sec:1}

Universality, according to which the same critical exponents occur
in all second-order phase transitions between the same two phases,
erstwhile phenomenologically established, has been a leading
principle of critical phenomena~\cite{Stanley}. The explanation of
universality, in terms of diverse Hamiltonian flows to a single
fixed point, has been one of the crowning achievements of
renormalization-group theory~\cite{Wilson}. In rather specialized
models in spatial dimension $d=2$, such as the
eight-vertex~\cite{Baxter} and Ashkin-Teller~\cite{Ashkin-Teller}
models, the critical exponents nevertheless vary continuously
along a line of second-order transitions, a phenomenon referred to
as the weak violation of universality. Although the existence and
quantitative description of universality classes in most pure spin
systems with simple interactions is well-established, this is not
true for more realistic models that include competing interactions
and/or disorder~\cite{cardy,young}.

In the current manuscript we discuss the universality aspects of
two distinct complex, in terms of randomness and interactions,
spin models, yielding exact information on their critical behavior
by applying finite-size scaling (FSS) techniques to high-accuracy
numerical data. In particular, we investigate the 2d Ising model
under the presence of quenched uncorrelated bond randomness and a
generalized 3-spin state Ising model with an additional
crystal-field coupling interaction in $d=3$, known as the the
Blume-Capel model~\cite{blume66,capel66}, for a certain regime of
its phase diagram. For the first 2d (random) model, for which its
exact phase diagram is known, we present comparative results for
two distinct values of the disorder strength and we give concrete
evidence in favor of the theoretically proposed strong
universality
hypothesis~\cite{DD-81,jug-83,shalaev-84,shankar-87,ludwig-87}.
For the latter 3d (pure) Blume-Capel model, which we choose to
simulate in the regime of its phase diagram where Ising-like
continuous transitions are known to take
place~\cite{blume66,capel66}, we consider a particular value of
the crystal field and present high-accuracy results for very large
lattice sizes that indeed place the model in the universality
class of the respective 3d Ising model.

Our study benefits from the Wang-Landau (WL)
algorithm~\cite{WL-01}, including some recently proposed
variations, namely the critical minimum energy subspace technique
of Malakis et al.~\cite{malakis-04} on the reduction of the energy
spectrum of the simulation, and the proposal of Belardinelli and
Pereyra~\cite{BP-07} regarding the application of the
energy-histogram flatness criterion of the original WL method. The
details of this implementation, as well as the general framework
of the WL approach are given in the next Section. Subsequently, in
Section~\ref{sec:3} we discuss the universality aspects of the
models, providing also estimates of their critical exponents,
which are found to be in good agreement with relevant existing
estimates in the literature. This paper is ended in
Section~\ref{sec:4} with a summary of our conclusions.

\section{Simulation method}
\label{sec:2}

Importance sampling methods have been for many years the main
tools in condensed matter physics and critical
phenomena~\cite{metropolis-53,bortz-75,binder-97,newman-99,landau-00}.
However, for complex systems, effective potentials may have a
rugged landscape, that becomes more pronounced with increasing
system size. In such cases, these traditional methods become
inefficient, since they cannot overcome large barriers in the
state space. A vast number of generalized ensemble methods have
been proposed to overcome this type of
problems~\cite{WL-01,newman-99,landau-00,lee-93,lee-06,oliveira-96,wang-99,berg-92,smith-95,torrie-97,swendsen-86,geyer-91,marinari-92,lyubartsev-92,hukushima-96,marinari-98,trebst-04}.
One important class of these methods emphasizes the idea of
directly sampling the energy density of states (DOS) and may be
called entropic sampling methods~\cite{newman-99}. In entropic
sampling, instead of sampling microstates with probability
proportional to $e^{-\beta E}$, one samples microstates with
probability proportional to $[G(E)]^{-1}$, where $G(E)$ is the
DOS, thus producing a flat energy histogram. The prerequisite for
the implementation of the method is the DOS information of the
system, a problem that can now be handled in many adequate ways
via a number of interesting approaches proposed in the last two
decades. The most remarkable examples are the Lee
entropic~\cite{lee-93,lee-06}, the
multicanonical~\cite{berg-92,smith-95}, the broad
histogram~\cite{oliveira-96}, the transition
matrix~\cite{wang-99}, the WL~\cite{WL-01}, and the optimal
ensemble methods~\cite{trebst-04}. In particular, there is a
considerable interest in the WL method and this is manifested in
the growing number of relevant publications that stem from nearly
every branch of the community of statistical mechanics. The WL
method is becoming a standard tool in examining simple, but also
more rugged, free-energy landscapes, in both magnetic and soft
matter systems and several papers dealing with improvements and
sophisticated implementations of its iterative process have
appeared throughout the
years~\cite{malakis-04,BP-07,douarche-03,troyer-03,fytas-06,schulz-05,reynal-05,jayasri-05,trebst-05,rathore-02,shell-02,yamaguchi-01,carri-05,calvo-03,tsai-07,vorontsov-04,poulain-06,schulz-03,dayal-04,fytas-08,alder-04,zhou-05,parsons-06,netto-06,antypov-08,wust-08,seaton-09,taylor-09,fytas11,wang-11}.

To apply the WL algorithm, an appropriate energy range of interest
has to be identified and a WL random walk is performed in this
energy subspace. Trials from a state with energy $E_{\rm i}$ to a
spin state with energy $E_{\rm f}$ are accepted according to the
transition probability
\begin{equation}
\label{eq:1}p(E_{\rm i}\rightarrow E_{\rm
f})=\min\left[\frac{G(E_{\rm i})}{G(E_{\rm f})},1\right].
\end{equation}
During the WL process the DOS $G(E)$ is modified $[G(E)\rightarrow
fG(E)]$ after each trial by a modification factor $f>1$. In the WL
process ($j=1,2,\cdots,j_{\rm final}$) successive refinements of
the DOS are achieved by decreasing the modification factor $f_{\rm
j}$. Most implementations use an initial modification factor
$f_{\rm j=1}=e\approx 2.71828\cdots$, a rule $f_{\rm
j+1}=\sqrt{f_{\rm j}}$, and a $5\%-10\%$ flatness criterion (on
the energy histogram) in order to move to the next refinement
level ($j\rightarrow j+1$)~\cite{WL-01}. The process is terminated
in a sufficiently high-level ($f\approx 1$, whereas the detailed
balanced condition limit is $f\rightarrow 1$).

In the last few years we have used an entropic sampling
implementation of the WL algorithm~\cite{WL-01} to study some
simple~\cite{malakis-04}, but also some more complex
systems~\cite{fytas-08}. One basic ingredient of this
implementation is a suitable restriction of the energy subspace
for the implementation of the WL algorithm. This was originally
termed as the critical minimum energy subspace
restriction~\cite{malakis-04} and it can be carried out in many
alternative ways, the simplest being that of observing the
finite-size behavior of the tails of the energy probability
density function of the system~\cite{malakis-04}. Assume that
$\tilde{E}$ denotes the value of energy producing the maximum term
in the partition function of the statistical model, at some
temperature of interest. Since we deal with a finite system of
linear size $L$, we are interested in the properties (finite-size
anomalies) near some pseudocritical temperature $T_{\rm
L}^{\ast}$, which in general depend on $L$ but also on the
property studied. Thus, we define a set of approximations by
restricting the statistical sums to energy subranges around the
value $\tilde{E}=\tilde{E}(T_{\rm L}^{\ast})$. Let these subranges
of the total energy range $(E_{\rm min},E_{\rm max})$ be denoted
as
\begin{equation}
\label{eq:2} (\tilde{E}_{-},\tilde{E}_{+}),\;\;\;\;
\tilde{E}_{\pm}=\tilde{E}\pm \Delta^{\pm},\;\;\;\;\;
\Delta^{\pm}\geq0.
\end{equation}
Accordingly
\begin{equation}
\label{eq:3} \tilde{\Phi}(E)=[S(E)-\beta
E]-\left[S(\tilde{E})-\beta
\tilde{E}\right],\;\;\tilde{Z}=\sum_{\tilde{E}_{-}}^{\tilde{E}_{+}}\exp{[\tilde{\Phi}(E)]}.
\end{equation}
Since by definition $\tilde{\Phi}(E)$ is negative we can easily
see that for large lattices extreme values of energy (far from
$\tilde{E}$) will have an extremely small contribution to the
statistical sums, since these terms decrease exponentially fast
with the distance from $\tilde{E}$. It follows that, if we request
a specified accuracy, then we may restrict the necessary energy
range in which DOS should be sampled. A simple idea is to use a
condition based on the energy probability density ($f_{T_{\rm
L}^{\ast}}(E)\propto \tilde{\Phi}(E)$), meaning the application of
Eq.~(\ref{eq:4}) at a particular pseudocritical temperature
$T_{\rm L}^{\ast}$. That is, we may define the end-points
($\tilde{E}_{\pm}$) of the subspaces by simply comparing the
corresponding probability densities with the maximum at the energy
$\tilde{E}$:
\begin{equation}
\label{eq:4} \tilde{E}_{\pm}:\;\;\;
\exp{\{\tilde{\Phi}(\tilde{E}_{\pm})\}}\leq r,
\end{equation}
where $r$ measures the relative error and it is usually set equal
to a small number ($r=10^{-6}$)~\cite{malakis-04}. This procedure
ends by providing us with a restricted energy subrange
$(E_{1},E_{2})$ centered around the value $\tilde{E}$, where the
large part of the simulation is finally carried out. Note that, in
general, the location of these subspaces can be predicted either
by extrapolation, from smaller lattices, or by using the
early-stage DOS approximation of the WL method.

Complications that may arise in random systems can be easily
accounted for by various simple modifications that take into
account possible oscillations in the energy probability density
function and expected sample-to-sample fluctuations of individual
disorder realizations. In our recent papers~\cite{fytas-08}, we
have presented details of various sophisticated routes for the
identification of the appropriate energy subspace $(E_{1},E_{2})$
for the entropic sampling of each random realization. In
estimating the appropriate subspace from a chosen pseudocritical
temperature one should be careful to account for the shift
behavior of other important pseudocritical temperatures and extend
the subspace appropriately from both low- and high-energy sides in
order to achieve an accurate estimation of all finite-size
anomalies. Of course, taking the union of the corresponding
subspaces, insures accuracy for the temperature region of all
studied pseudocritical temperatures.

The up to date version of our implementation uses a combination of
several stages of the WL process. First, we carry out a starting
(or preliminary) multi-range (multi-R) stage, in a very wide
energy subspace. This preliminary stage may consist of the levels:
$j=1,\cdots,18$ of the adjustment of the modification factor and
to improve accuracy the process may be repeated several times.
However, in repeating the preliminary process and in order to be
efficient, we use only the levels $j=13,\cdots,18$ after the first
attempt, using as starting DOS the one obtained in the first
random walk at the level $j=12$. From our experience, this
practice is almost equivalent of simulating the same number of
independent WL random walks. Also in our recent studies we have
found out that is much more efficient and accurate to loosen up
the originally applied very strict flatness
criteria~\cite{WL-01,malakis-04}. Thus, a variable flatness
process starting at the first levels with a very loose flatness
criteria and assuming at the level $j=18$ the original strict
flatness criteria is nowadays used. After the above described
preliminary multi-R stage, in the wide energy subspace, one can
proceed in a safe identification of the appropriate energy
subspace using one or more alternatives outlined in
Ref.~\cite{malakis-04}. In random systems, where one needs to
simulate many disorder realizations, it is also possible and
advisable to avoid the identification of the appropriate energy
subspace separately for each disorder realization by extrapolating
from smaller lattices and/or by prediction from preliminary runs
on small numbers of disorder realizations. In any case, the
appropriate subspaces should be defined with sufficient
tolerances. In our implementation we use such advance information
to proceed in the next stages of the entropic sampling.

The process continues in two further stages (two-stage process),
using now mainly high iteration levels, where the modification
factor is very close to unity and there is not any significant
violation of the detailed balance condition during the WL process.
These two stages are suitable for the accumulation of energy and
magnetization ($E,M$) histograms, which can be used for an
accurate entropic calculation of non-thermal thermodynamic
parameters, such us the order parameter $M$ and its susceptibility
$\chi$~\cite{malakis-04}. In particular, the resulting
approximation of the DOS and the corresponding $E,M$ histograms
may be used to estimate the magnetic properties of the system in a
temperature range, which is covered, by the restricted energy
subspace $(E_{1},E_{2})$. Canonical averages of the form
\begin{equation}
\label{eq:5}\langle M^{n}\rangle=\frac{\sum_{E}\langle
M^{n}\rangle_{E}G(E)e^{-\beta E}}{\sum_{E}G(E)e^{-\beta E}},
\end{equation}
where $G(E)$ denotes the exact DOS, will be then approximated via
\begin{equation}
\label{eq:6} \langle M^{n}\rangle\cong
\frac{\sum_{E\in(E_{1},E_{2})}\langle M^{n}\rangle_{\rm
E,WL}G_{\rm WL}(E)e^{-\beta E}}{\sum_{E\in(E_{1},E_{2})}G_{\rm
WL}(E)e^{-\beta E}},
\end{equation}
with $G_{\rm WL}(E)$ the DOS of the above described WL process.
Then, the microcanonical averages $\langle M^{n}\rangle_{E}$ are
obtained from the $H_{\rm WL}(E,M)$ histograms through the
following formulae
\begin{equation}
\label{eq:7}\langle M^{\rm n}\rangle_{E}\cong\langle M^{\rm
n}\rangle_{\rm E,WL}\equiv \sum_{\rm M}M^{\rm n}\frac{H_{\rm
WL}(E,M)}{H_{\rm WL}(E)}.
\end{equation}
As usual $H_{\rm WL}(E)=\sum_{M}H_{\rm WL}(E,M)$ and the summation
in $M$ runs over all values generated during the process in the
restricted energy subspace $(E_{1},E_{2})$. The accuracy of the
magnetic properties obtained from the above averaging process will
depend on many factors. Firstly, the used energy subspace
restricts the temperature range for which such approximations may
be accurate. This restriction has as a result that the process
will not visit all possible values of $M$, but this fact is of no
consequence for the accuracy of the magnetic properties at the
temperature range of interest, as far as the estimated DOS is
accurate. Secondly, the accuracy of the above microcanonical
estimators will, as usually, depend on the total number of visits
to a given energy level $[H_{\rm WL}(E)]$, and also to the number
of different spin states visited within this energy level.
However, these are statistical fluctuations inherent in any Monte
Carlo method and we should expect improvement by increasing the
number of repetitions of the process.

In the first (high-level) stage, we follow again a repeated
several times (typically $\sim 5-10$) multi-R WL approach, carried
out now only in the restricted energy subspace. The WL levels may
be now chosen as $j=18,19,20$ and as an appropriate starting DOS
for the corresponding starting level the average DOS of the
preliminary stage at the starting level may be used. Finally, the
second (high-level) stage is applied in the refinement WL levels
$j=j_{\rm i},\cdots,j_{\rm i}+4$ (typically $j_{\rm i}=21$), where
we usually test both an one-range (one-R) or a multi-R approach
with large energy intervals. We should note here however that,
most authors use the more efficient multi-R approach in the final
stages of the WL process, as a consequence of the well-known slow
convergence of the method at the high iteration-levels. We point
out here that, it is possible to overcome this slow convergence by
using a looser flatness criterion or an alternative Lee entropic
final stage, as proposed in Ref.~\cite{lee-06} and applied in
Ref.~\cite{fytas-08}. Moreover, recently a different alternative
has been proposed by Belardinelli and Pereyra (denoted hereafter
as the BP approach)~\cite{BP-07}. Following their proposal, one is
using, in the final stage, an almost continuously changing
modification factor adjusted according to the rule $\ln{f}\sim
t^{-1}$. Since $t$ is the Monte Carlo time, using a time-step
conveniently defined proportional to the size of the energy
subinterval, the efficiency of this scheme is independent of the
size of the subintervals and therefore the method provides the
same efficiency in both multi-R and one-R approaches. Furthermore,
from the tests performed by these authors, and also from our
comparative studies in the 2d pure and random-bond Ising model, as
will be seen below in Section~\ref{sec:3}, the error-behavior of
this method seems superior to the original WL process, improving
to some extent the saturation-error problem of the WL method and
giving slightly better estimates of critical temperatures and
exponents. Accordingly, we have also applied this alternative
route for the final stage of our simulations using an one-R
approach.

The above described numerical approach was used to estimate the
critical properties of the considered spin models. In particular,
for the random-bond version of the 2d Ising model we have
simulated a very large number of disorder realizations (of the
order of $500 - 1000$, disorder averaging is symbolized as usual
with $[\cdots]_{\rm av}$) for both values of the disorder strength
considered and for systems with linear sizes in the range
$L=20-200$. For each lattice size we have performed two types of
simulations: the original multi-R WL approach and the WL approach
modified by the recent BP proposal~\cite{BP-07}. In both cases the
exact same disorder realizations have been generated and
simulated, in order to have a direct comparison of the extracted
critical behavior. Finally, for the pure version of the 3d
Blume-Capel model at $\Delta=0$, we simulated, using only the BP
procedure in the final stage of our combined algorithm, simple
cubic lattices with $N=L^{3}$ spins, where $L=8-64$. For this
case, each lattice size was simulated $100$ times with different
initial random numbers to get a better statistical analysis of the
results.

Before closing, let us comment on the nature of our error bars
illustrated in the following figures below and also used in the
corresponding fitting attempts. For the case of the random-bond
model, even for the larger lattice sizes studied, the statistical
errors of the WL method were found to be of reasonable magnitude
and in some cases to be of the order of the symbol sizes, or even
smaller. Thus in the figures below we choose to show only the
errors due to the finite number of disorder realizations. These
errors have been estimated by two similar methods, using groups of
$25$ to $50$ realizations for each lattice size and the jackknife
method or a straightforward variance calculation (blocking
method)~\cite{newman-99}. The jackknife method yielded some
reasonably conservative errors, about $10-20\%$ larger than the
corresponding calculated standard deviations, and are shown as
error bars in our figures. Finally, let us point out that in all
cases studied, the sample-to-sample fluctuations for the
individual maxima are much large than the corresponding finite
disorder sampling errors. Finally, for the case of the pure
Blume-Capel model, the error bars shown reflect the sample
variance of the $100$ independent runs performed in each lattice
size.

\section{Results and discussion}
\label{sec:3}

\subsection{Strong universality in the 2d random-bond Ising model}
\label{sec:3a}

Understanding the role played by impurities on the nature of phase
transitions is of great importance, both from experimental and
theoretical perspectives. First-order phase transitions are known
to be dramatically softened under the presence of quenched
randomness~\cite{aizenman-89,hui-89,berker-93,chen-92,cardy-96,cardy-97,chatelain-98,paredes-99,chatelain-01,fernandez-08},
while continuous transitions may have their exponents altered
under random fields or random bonds~\cite{harris-74,chayes-86}.
There are some very useful phenomenological arguments and some,
perturbative in nature, theoretical results, pertaining to the
occurrence and nature of phase transitions under the presence of
quenched
randomness~\cite{hui-89,cardy-96,dotsenko-95,jacobsen-98}.
Historically, the most celebrated criterion is that suggested by
Harris~\cite{harris-74}. This criterion relates directly the
persistence, under random bonds, of the non random behavior to the
specific heat exponent $\alpha_{\rm p}$ of the pure system.
According to this criterion, if $\alpha_{\rm p}>0$, then disorder
will be relevant, i.e., under the effect of the disorder, the
system will reach a new critical behavior. Otherwise, if
$\alpha_{\rm p}<0$, disorder is irrelevant and the critical
behavior will not change.

\begin{figure}
\resizebox{1 \columnwidth}{!}{\includegraphics{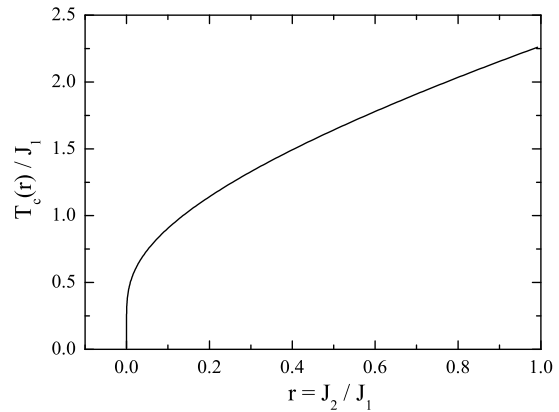}}
\caption{Critical temperatures as a function of the disorder
strength $r$ of the 2d random-bond Ising model defined in
Eqs.~(\ref{eq:8}) and (\ref{eq:9}) obtained by solving
Eq.~(\ref{eq:10}).} \label{fig:1}
\end{figure}

Pure systems with a zero specific heat exponent are marginal cases
of the Harris criterion and their study, upon the introduction of
disorder, has been of particular interest~\cite{MK-99}. The
paradigmatic model of the marginal case is the general random 2d
Ising model (random-site, random-bond, and bond-diluted) and this
model has been extensively investigated and debated [see
Refs.~\cite{kenna,gordillo-09} and references therein]. Several
recent studies, both analytical (renormalization group and
conformal field theories) and numerical (mainly Monte Carlo
simulations) devoted to this model, have provided very strong
evidence in favor of the so-called logarithmic corrections's
scenario~\cite{DD-81,jug-83,shalaev-84,shankar-87,ludwig-87,wang-90}.
According to this, the effect of infinitesimal disorder gives rise
to a marginal irrelevance of randomness and besides logarithmic
corrections, the critical exponents maintain their 2d Ising
values. Here, we should mention that there is not full agreement
in the literature and a different scenario, the so-called weak
universality scenario~\cite{KP-94,kuhn-94,suzuki-74,gunton-75},
predicts that critical quantities, such as correlation length
display power-law singularities, with the corresponding exponents
$\nu$ changing continuously with the disorder strength; however
this variation is such that the magnetic exponent ratios remain
constant at the pure system's value.

\begin{figure}
\resizebox{1 \columnwidth}{!}{\includegraphics{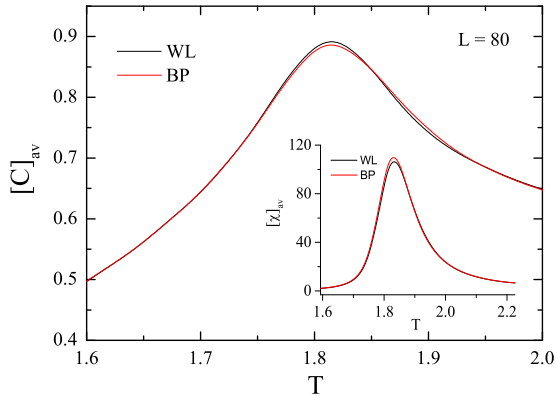}}
\caption{(color online) A comparative plot of the
disorder-averaged specific heat (main panel) and magnetic
susceptibility (inset) as a function of temperature for a lattice
size $L=80$ and disorder strength $r=r_{1}=1/7$ of the 2d
random-bond Ising model.} \label{fig:2}
\end{figure}

Here we present concrete evidence in favor of the strong
universality scenario together with a comparative algorithmic test
of the WL-type of methods presented above. Let us at this point
define the quenched version of the 2d Ising model that we employ.
In particular we choose to study here the random-bond version of
the square lattice Ising model (RBIM), which is defined with the
help of the following Hamiltonian
\begin{equation}
\label{eq:8} \mathcal{H}^{\rm
(RBIM)}=-\sum_{<ij>}J_{ij}s_{i}s_{j}.
\end{equation}
In the above Eq.~(\ref{eq:8}) the spin variables $s_{\rm i}$ take
on the values $-1,+1$, $<ij>$ indicates summation over all
nearest-neighbor pairs of sites, and $J_{\rm ij}$ is the
ferromagnetic exchange interaction taken from a bimodal, quenched
bond-disorder, distribution of the form
\begin{equation}
\label{eq:9}
\mathcal{P}(J_{ij})=\frac{1}{2}~[\delta(J_{ij}-J_{1})+\delta(J_{ij}-J_{2})],
\end{equation}
where $J_{1}+J_{2}=2$, $J_{1}>J_{2}>0$, and $r=J_{2}/J_{1}$
reflects the strength of the bond randomness. We also fix $2k_{\rm
B}/(J_{1}+J_{2})=1$ to set the temperature scale.
\begin{figure}
\resizebox{1 \columnwidth}{!}{\includegraphics{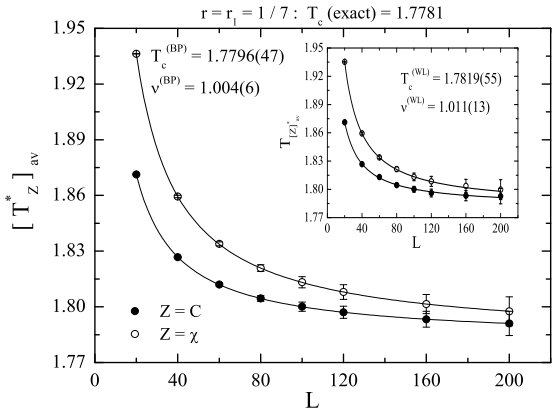}}
\caption{Shift behavior of the pseudocritical temperatures of the
specific heat and susceptibility, averaged over disorder, for the
2d random-bond Ising model with disorder strength $r=r_{1}=1/7$.
The solid lines show a simultaneous fitting according to the power
law described in the text for the total lattice range $L=20-200$.
The main panel shows the data and results obtained via the BP
approach, whereas the inset via the WL approach.} \label{fig:3}
\end{figure}
With the above distribution the 2d random model exhibits a unique
advantage, that is its critical temperature $T_{\rm c}$ is exactly
known~\cite{fisch-78} as a function of the disorder strength $r$
through the relation
\begin{equation}
\label{eq:10} \sinh{(2J_{1}/T_{\rm c})}\cdot \sinh{(2rJ_{1}/T_{\rm
c})}=1.
\end{equation}
The solution of the above Eq.~(\ref{eq:10}), i.e. the function
$T_{\rm c}(r)$, is plotted in Fig.~\ref{fig:1} and consists the
phase diagram of the model in the temperature - disorder strength
plane.

This unique feature of knowing exactly the critical temperature of
a disordered model gives us the opportunity to perform a
comparative FSS analysis of several pseudocritical temperatures
for extracting estimates of the critical temperatures and shift
exponents using the algorithmic procedures described above. In our
simulations we have considered two values of the disorder
strength, namely the values $r=r_{1}=1/7$ and $r=r_{2}=1/9$, both
belonging to the strong-disorder regime of the model. The exact
critical temperatures for these values of $r$ are $T_{\rm
c}(r=r_{1})=1.7781\cdots$ and $T_{\rm c}(r=r_{2})=1.6853\cdots$,
respectively.

\begin{figure}
\resizebox{1 \columnwidth}{!}{\includegraphics{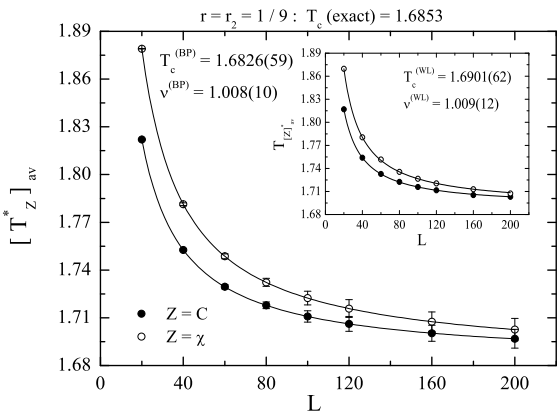}}
\caption{Similar to Fig.~\ref{fig:3} but for a different value of
the disorder strength, namely $r=r_{2}=1/9$.} \label{fig:4}
\end{figure}

Our results for the 2d random-bond Ising model are shown in
Figs.~\ref{fig:2} - \ref{fig:4}. We start with Fig.~\ref{fig:2}
which is a comparative illustration of the disorder-averaged
specific heat $[C]_{\rm av}$ and susceptibility $[\chi]_{\rm av}$
curves as a function of temperature, estimated via the two
approaches, WL and BP, used in this study. One may observe some
small differences in the location (and the corresponding maximum
value) of the peak in both specific heat and susceptibility data
via the two implemented approaches. Subsequently, in
Figs.~\ref{fig:3} and \ref{fig:4} we plot the shift behavior of
the disorder averaged pseudocritical temperatures of the specific
heat and magnetic susceptibility as a function of the lattice size
$L$ and in all cases the solid lines illustrate a simultaneous
fitting procedure of the form $[T_{\rm Z}^{\ast}]_{\rm av}=T_{\rm
c}+b_{\rm z} L^{-1/\nu}$, where $Z=C$ or $Z=\chi$, as shown by the
corresponding symbols (filled and open circles) in the figures. In
the main panels we present our numerical data obtained by applying
in the final stage of the algorithmic procedure the BP approach
(thus denoted by the superscript (BP) in the figures), whereas in
the corresponding insets those of the original multi-R WL process
(thus denoted by the superscript (WL) in the figures).

It is clear from both figures that the results obtained for the
critical temperatures and also for the critical exponent of the
correlation length are in excellent agreement with the exact
values and also support the strong universality scenario
hypothesis, via the relation $\nu=1$. Also, they compare well to
the results given by the extensive Swendsen-Wang Monte Carlo
analysis of Wang et al.~\cite{wang-90}. Finally, let us comment
here that, although the computational time used for the BP
approach is somewhat larger (of the order of $\sim 2.5$) than that
of the simple multi-R WL process, the estimates obtained via this
approach are slightly better than those of the simple multi-R WL
procedure and this is true for both values of the disorder
strength considered in this paper. Furthermore, other relevant
tests performed originally in the simple case of the pure 2d Ising
model and our experience of simulating rough random systems, such
as the random-field Ising model~\cite{fytas-08} or systems
undergoing first-order phase transitions in their pure
versions~\cite{malakis09}, support the above illustration.

\subsection{Ising universality in the Blume-Capel model}
\label{sec:3b}

We investigate the universality of the 3d Blume-Capel in the
second-order phase-transition regime of its phase diagram
(temperature - crystal field plane) using the algorithmic
procedure that combines a repetitive application of the WL
algorithm in the first stage and the BP approach in the final
stage of the algorithm. For this purpose, the static critical
exponents are estimated by analyzing the obtained numerical data
within the framework of the well-established FSS theory. At the
same time, their values are calculated using power-law relations
of related thermodynamic quantities. As also stated below, the 3d
Blume-Capel model is expected to be in the universality class of
the 3d Ising model for the second-order phase-transition regime of
its phase diagram. Thus, it is crucial at this point to remind the
reader some of the best well-known estimates in the literature for
the critical exponents of the 3d Ising model, as given in
Ref.~\cite{guida-98}: $\nu=0.6304(13)$, $\gamma/\nu=1.966(3)$, and
$\beta/\nu=0.517(3)$. For the most accurate complete set of
critical exponents to the 3d Ising universality class we refer the
reader to the review by of Pelissetto and Vicari~\cite{peli}.

Let us define at this point the Hamiltonian of the Blume-Capel
(BC) model~\cite{blume66,capel66}
\begin{equation}
\label{eq:11} \mathcal{H}^{\rm
(BC)}=-J\sum_{<ij>}s_{i}s_{j}+\Delta\sum_{i}s_{i}^{2},
\end{equation}
where the spin variables $s_{\rm i}$ take on the values $-1, 0$,
or $+1$, as usual $<ij>$ indicates summation over all
nearest-neighbor pairs of sites, and $J>0$ is the ferromagnetic
exchange interaction. The parameter $\Delta$ is known as the
crystal-field coupling and to fix the temperature scale we set
$J=1$ and $k_{\rm B}=1$. This model is of great importance for the
theory of phase transitions and critical phenomena and besides the
original mean-field theory~\cite{blume66,capel66}, has been
analyzed by a variety of approximations and numerical approaches,
in both 2d and 3d. These include the real space renormalization
group, Monte Carlo simulations, and Monte Carlo
renormalization-group calculations~\cite{landau72},
$\epsilon$-expansion renormalization groups~\cite{stephen73},
high- and low-temperature series calculations~\cite{fox73}, a
phenomenological FSS analysis using a strip
geometry~\cite{nightingale82,beale86}, and Monte Carlo
simulations~\cite{malakis09,jain80,landau81,care93,deserno97,blote03,silva06,kutlu06,hasen10}.
The phase diagram of the model consists of a segment of continuous
Ising-like transitions at high temperatures and low values of the
crystal field which ends at a tricritical point, where it is
joined with a second segment of first-order transitions between
($\Delta_{\rm t},T_{\rm t}$) and ($\Delta_{0},T=0$). For the
simple cubic lattice, considered in this paper, $\Delta_{0}=3$.
The location of the tricritical point has been estimated by
Deserno~\cite{deserno97}, via a microcanonical Monte Carlo
approach, and is given by $[\Delta_{\rm t},T_{\rm
t}]=[2.84479(30),1.4182(55)]$.

\begin{figure}
\resizebox{1 \columnwidth}{!}{\includegraphics{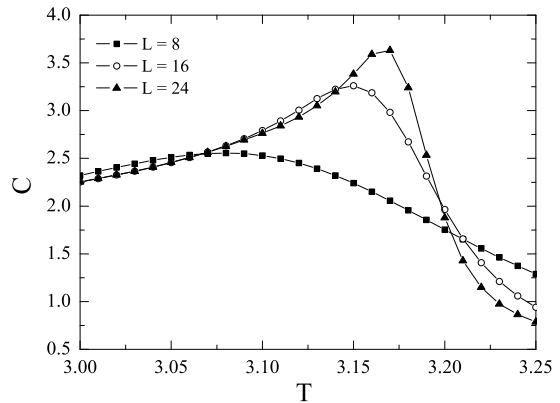}}
\caption{Specific heat curves of the 3d $\Delta=0$ Blume-Capel
model as a function of temperature for $L=8$, $16$, and $L=24$.}
\label{fig:5}
\end{figure}

\begin{figure}
\resizebox{1 \columnwidth}{!}{\includegraphics{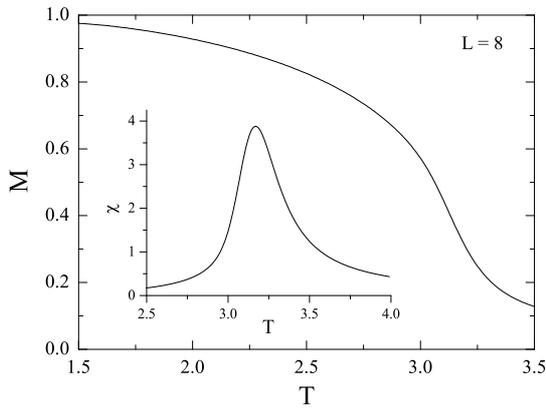}}
\caption{Order-parameter (main panel) and corresponding magnetic
susceptibility (inset) of the 3d $\Delta=0$ Blume-Capel model as a
function of the temperature for a lattice with linear size $L=8$.}
\label{fig:6}
\end{figure}

\begin{figure}
\resizebox{1 \columnwidth}{!}{\includegraphics{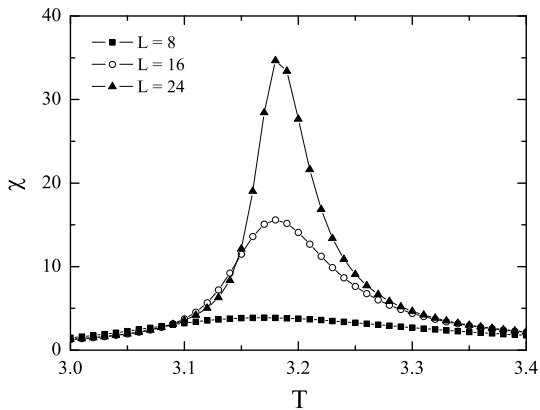}}
\caption{Magnetic susceptibility curves of the 3d $\Delta=0$
Blume-Capel model as a function of temperature for $L=8$, $16$,
and $L=24$.} \label{fig:7}
\end{figure}

In what follows we restrict our analysis to the value $\Delta=0$
and we simulate the Blume-Capel model for a wide range of
lattices, with linear sizes $L\in \{8,16,24,32,48,64\}$, summing
for each lattice size $100$ independent simulations with different
initial conditions in order to gain better statistics. Moreover,
in all fitting procedures shown in Figs.~\ref{fig:8} -
\ref{fig:10} below, we take account the complete lattice range. We
should note here that, recent Monte Carlo simulations of the model
have been restricted to lattice sizes of the order of
$L=24$~\cite{kutlu06}. The first set of our illustrations,
Figs.~\ref{fig:5} - \ref{fig:7}, is rather instructive. We show
various thermodynamic quantities for typical lattice sizes used in
the scaling analysis below. In particular, in Fig.~\ref{fig:5} we
plot the specific heat of the model as a function of the
temperature for three characteristic linear sizes, i.e., $L=8$,
$16$, and $L=32$. One can observe from this figure the clear shift
of the pseudocritical temperatures with increasing lattice size,
as well as the increase in the value of the corresponding maximum.
Figure~\ref{fig:6} now shows for a lattice with linear size $L=8$
the temperature behavior of the order parameter $M$ in the main
panel, and its corresponding deduced magnetic susceptibility
$\chi$ in the inset. Finally, Fig.~\ref{fig:7} shows the analogous
$L$ - dependence behavior of the susceptibility curves as a
function of the temperature, where again, as in Fig.~\ref{fig:5}
the expected shift behavior is recovered.

\begin{figure}
\resizebox{1 \columnwidth}{!}{\includegraphics{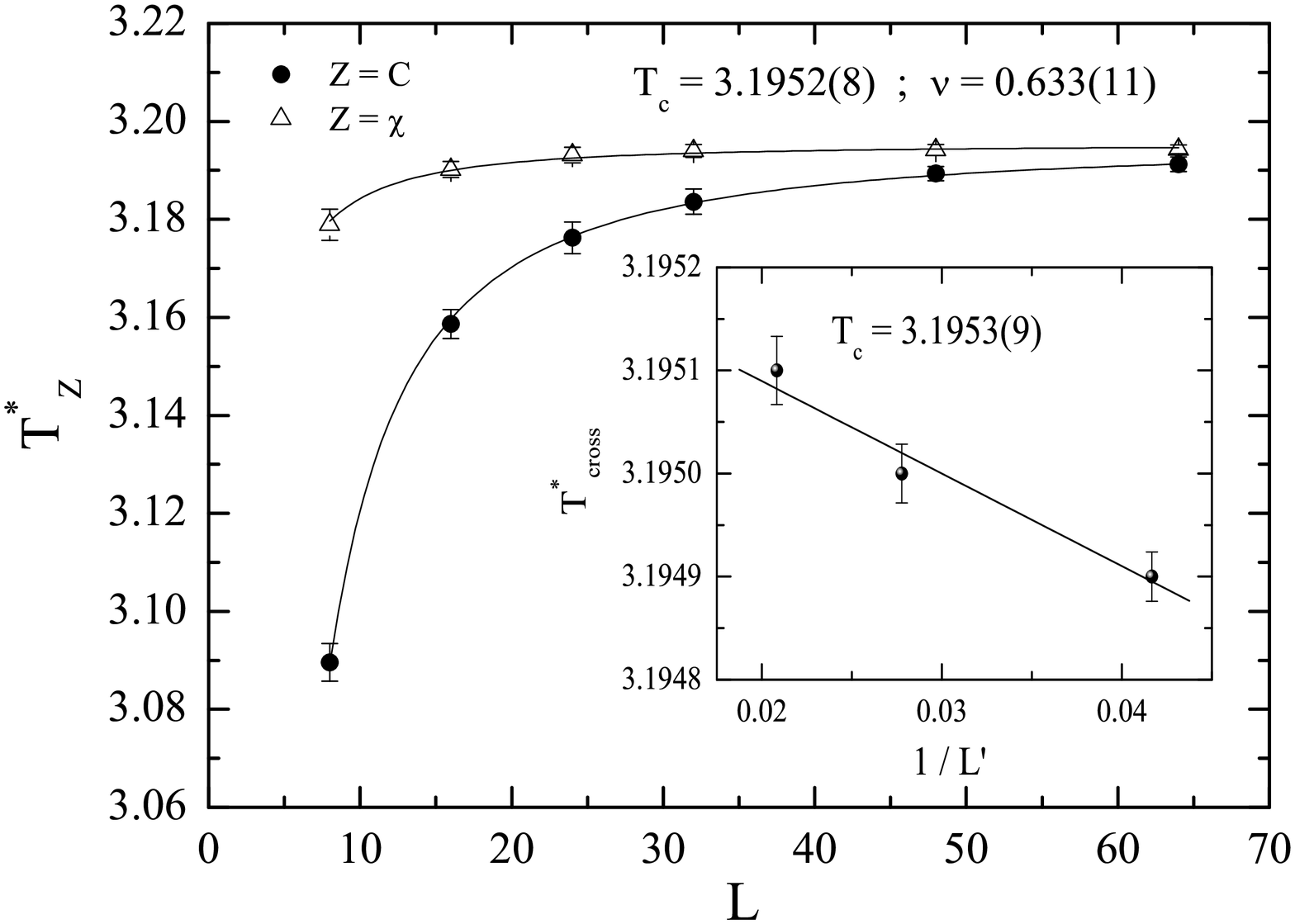}}
\caption{Shift behavior of the pseudocritical temperatures of the
specific heat and susceptibility for the 3d Blume-Capel model at
the value $\Delta=0$. The inset shows the FSS behavior of the
crossings of the fourth-order Binder's cumulant with inverse
lattice size (see also discussion in the text).} \label{fig:8}
\end{figure}

We proceed now with the FSS analysis of our numerical data in
Figs.~\ref{fig:8} - \ref{fig:10}. In particular, in the main panel
of Fig.~\ref{fig:8} we present the simultaneous fitting of two
pseudocritical temperatures of the model, namely those of the
specific heat $C$ and magnetic susceptibility $\chi$, according to
the well-known power law $T_{\rm Z}^{\ast}=T_{\rm c}+b_{\rm z}
L^{-1/\nu}$. The result we get for the critical temperature, also
shown in the panel, $T_{\rm c}=3.1952(8)$, is in very good
agreement with the Ornstein-Zernike approximation of the phase
diagram of the model by Grollau \emph{et al.}~\cite{grollau01} and
with the numerical estimation $3.20(1)$ given in
Ref.~\cite{kutlu06}. Additionally, the estimated value of the
critical exponent $\nu$, $\nu=0.633(11)$, is in excellent
agreement with the value $0.6304(13)$ of the simple 3d Ising
model~\cite{guida-98} and the value $0.63002(10)$ quoted by the
most accurate calculation of Hasenbusch for the second-order
phase-transition regime of the Blume-Capel model~\cite{hasen10}.
This is a clear first strong indication of the Ising universality
in the second-order regime of the Blume-Capel model. In the
corresponding inset of Fig.~\ref{fig:8} we illustrate a further
test of our accuracy of the WL and BP approaches, by using the
crossings of the fourth-order Binder's cumulant $U_{\rm
L}=1-\langle M^4\rangle/ \left[3\langle
M^2\rangle^2\right]$~\cite{landau-00}. We show three data points
which denote the temperature crossing points ($T^{\ast}_{\rm
cross}$) of the Binder's cumulant for the following pairs of
lattices: $(L_{1},L_{2})=(16,32)$, $(24,48)$, and $(32,64)$. The
notation $L'$ in the x-axis refers to the value
$L'=(L_{1}+L_{2})/2$. The solid line is a simple linear fitting
extrapolating to $L'\rightarrow \infty$, which gives an estimate
for the critical temperature $T_{\rm c}=3.1953(9)$, also in very
good agreement with the estimate of the main panel.

\begin{figure}
\resizebox{1 \columnwidth}{!}{\includegraphics{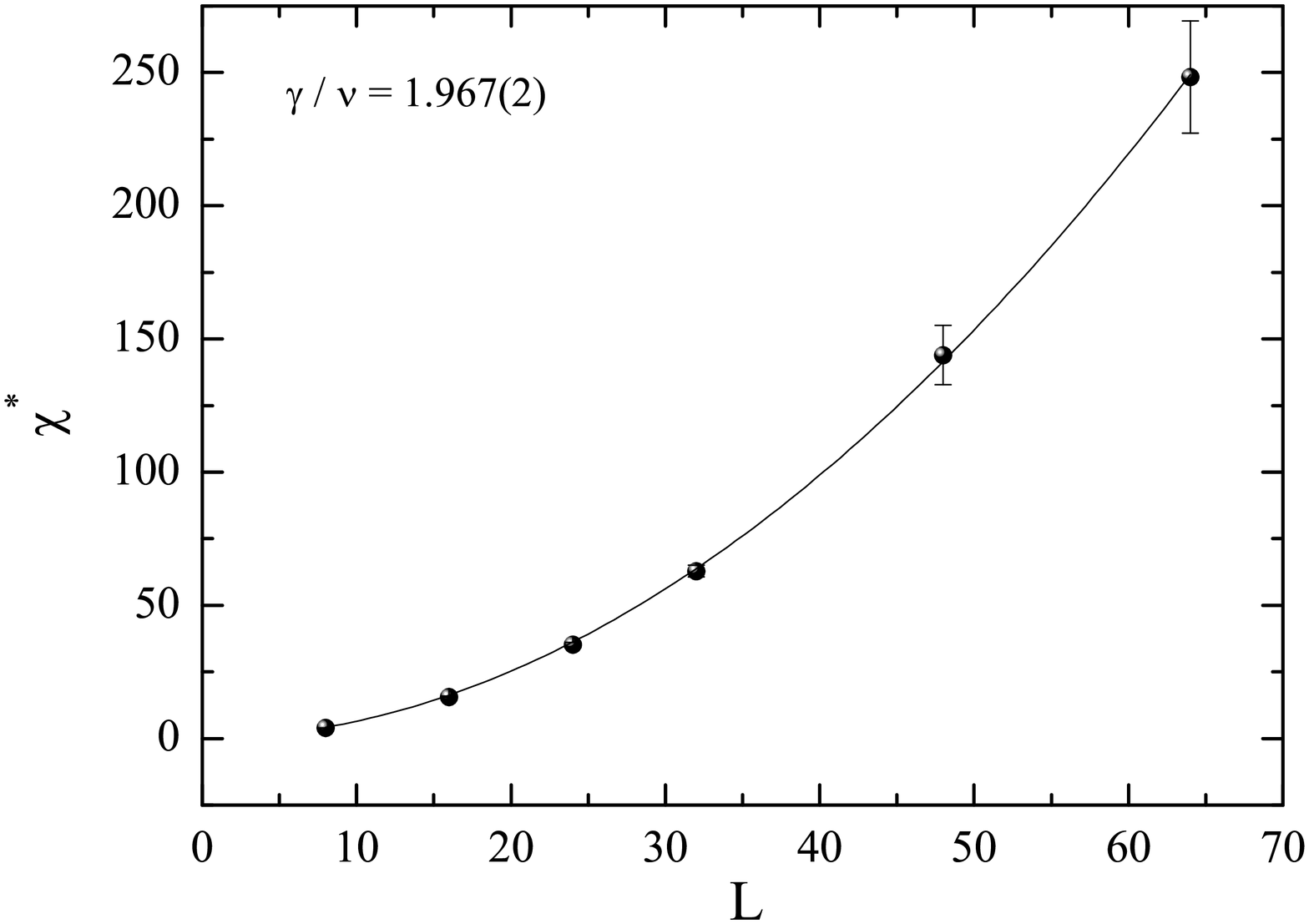}}
\caption{FSS behavior of the magnetic susceptibility maxima of the
3d $\Delta=0$ Blume-Capel model.} \label{fig:9}
\end{figure}

We proceed with the estimation of the magnetic exponent ratios of
the 3d Blume-Capel model at $\Delta=0$. Figure~\ref{fig:9}
illustrates the FSS behavior of the magnetic susceptibility maxima
which are expected to scale as $\chi^{\ast}\sim L^{\gamma/\nu}$
with the lattice size. The solid line is a fitting of the above
form giving an estimate of $1.967(2)$ for the exponent ratio
$\gamma/\nu$, very close to the value $1.966(3)$ of the simple 3d
Ising model~\cite{guida-98}. In Fig.~\ref{fig:10} we present
numerical data for the order parameter $M$ of the model at the
estimated critical temperature of Fig.~\ref{fig:8}. The solid line
is simple power-law fitting of the form $M_{\rm c}\sim
L^{-\beta/\nu}$ which gives the estimate $0.518(4)$ for the
critical exponent ratio, very close to the value $0.517(3)$ of the
corresponding pure Ising model~\cite{guida-98}. We note here that,
both of these magnetic exponent ratios are a considerable
improvement of the recent numerical estimations of \"{O}zkan et
al.~\cite{kutlu06} using two different procedures defined as the
standard and cooling algorithms on a cellular automaton estimation
that gave the values $1.94(4)$ and $0.48(6)$ for the exponent
ratios $\gamma/\nu$ and $\beta/\nu$, respectively.

\begin{figure}
\resizebox{1 \columnwidth}{!}{\includegraphics{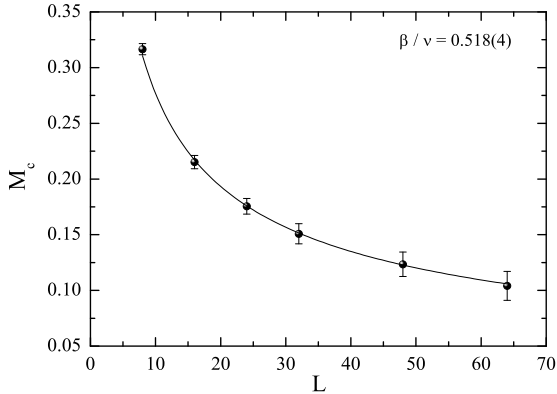}}
\caption{FSS behavior of the critical order parameter data of the
3d $\Delta=0$ Blume-Capel model.} \label{fig:10}
\end{figure}

Overall, the FSS analysis performed in this Section places,
without doubt, the second-order phase-transition regime of
Blume-Capel model to the universality class of the pure 3d Ising
model. Additionally, the proposed estimates for the critical
exponents clearly indicate the accuracy of the numerical scheme
that is based on the WL algorithm.

\section{Summary and outlook}
\label{sec:4}

Summarizing, in the present paper we reported large-scale
numerical simulations of the random 2d Ising model and the pure 3d
Blume-Capel model using a modified version of the Wang-Landau
algorithm. For the first (random) model, both sets of the
numerical data and the relevant finite-size scaling analysis for
the two values of the disorder strength considered, clearly
support the strong universality scenario hypothesis, through the
estimated value of the correlation length's exponent $\nu\cong 1$.
Furthermore, the accuracy of the simulation scheme via the
Wang-Landau algorithm has been clearly shown via a direct
comparison between the estimated values of the critical
temperatures and the exact ones. For the latter (pure) model, our
estimates of the critical exponents compare well with the most
accurate ones in the current literature and place the second-order
phase-transition regime of the Blume-Capel model to the
universality class of the pure 3d Ising model, as also expected on
theoretical grounds. Further attempts are currently being
considered in order to obtain a complete numerical derivation of
the Blume-Capel model's phase diagram by simulating several values
of the crystal field $\Delta$, including also those in the model's
first-order phase-transition regime.

\begin{acknowledgement}
N.G.F. has been partly supported by MICINN, Spain, through
Research Contract No. FIS2009-12648-C03. P.E.T. is grateful for
financial support by the Austrian Science Foundation within the
SFB ViCoM (Grant F41).
\end{acknowledgement}

{}


\begin{thebibliography}{}
\bibitem{Stanley} H.E. Stanley, Introduction to Phase Transitions and Critical Phenomena (Oxford U.P., Oxford, 1971)
\bibitem{Wilson} K.G. Wilson, Phys. Rev. B {\bf 4}, 3174 (1971)
\bibitem{Baxter} R.J. Baxter, Ann. Phys. {\bf 70}, 193 (1970)
\bibitem{Ashkin-Teller} J. Ashkin, E. Teller, Phys. Rev. {\bf 64}, 178 (1943).
\bibitem{cardy} J. Cardy, {\em Scaling and Renormalization in Statistical
Physics}, (Cambridge University Press 1996)
\bibitem{young} A.P. Young (ed.), {\em Spin glasses and random fields},
(World Scientific, Singapore 1998)
\bibitem{blume66} M. Blume, Phys. Rev. {\bf 141}, 517 (1966)
\bibitem{capel66} H.W. Capel, Physica (Utr.) {\bf 32}, 966 (1966); H.W. Capel, Physica (Utr.)
{\bf 33}, 295 (1967); H.W. Capel, Physica (Utr.) {\bf 37}, 423 (1967)
\bibitem{DD-81} Vik.S. Dotsenko, Vl.S. Dotsenko, JETP Lett. {\bf 33}, 37 (1981)
\bibitem{jug-83} G. Jug, Phys. Rev. B {\bf 27}, 4518 (1983)
\bibitem{shalaev-84} B.N. Shalaev, Sov. Phys. Solid State {\bf 26}, 1811 (1984)
\bibitem{shankar-87} R. Shankar, Phys. Rev. Lett. {\bf 58}, 2466 (1987)
\bibitem{ludwig-87} A.W.W. Ludwig, Nucl. Phys. B {\bf 285}, 97 (1987)
\bibitem{WL-01} F. Wang, D.P. Landau, Phys. Rev. Lett. {\bf 86}, 2050 (2001); F. Wang, D.P. Landau, Phys. Rev. E {\bf 64}, 056101 (2001)
\bibitem{malakis-04} A. Malakis, A. Peratzakis, N.G. Fytas, Phys. Rev. E {\bf
70}, 066128 (2004); A. Malakis, S.S. Martinos, I.A. Hadjiagapiou,
N.G. Fytas, P. Kalozoumis, Phys. Rev. E {\bf 72} 066120 (2005)
\bibitem{BP-07} R.E. Belardinelli, V.D. Pereyra, Phys. Rev. E {\bf 75}, 046701
(2007); R.E. Belardinelli, V.D. Pereyra, J. Chem. Phys. {\bf 127},
184105 (2007)
\bibitem{metropolis-53} N. Metropolis, A.W. Rosenbluth, M.N. Rosenbluth, A.H. Teller, J. Chem. Phys. {\bf 21}, 1087 (1953)
\bibitem{bortz-75} A.B. Bortz, M.H. Kalos, J.L. Lebowitz, J. Comput. Phys. {\bf 17}, 10 (1975)
\bibitem{binder-97} K. Binder, Rep. Prog. Phys. {\bf 60}, 487 (1997)
\bibitem{newman-99} M.E.J. Newman, G.T. Barkema, \textit{Monte Carlo Methods in
Statistical Physics} (Clarendon Press, Oxford, 1999)
\bibitem{landau-00} D.P. Landau, K. Binder, \textit{A Guide to Monte Carlo Simulations in
Statistical Physics} (Cambridge University Press, Cambridge, 2000)
\bibitem{lee-93} J. Lee, Phys. Rev. Lett. {\bf 71}, 211 (1993)
\bibitem{lee-06} H.K. Lee, Y. Okabe, D.P. Landau, Comput. Phys. Commun. {\bf 175}, 36 (2006)
\bibitem{oliveira-96} P.M.C. de Oliveira, T.J.P. Penna, H.J. Herrmann, Braz. J. Phys. {\bf 26}, 677 (1996)
\bibitem{wang-99} J.-S. Wang, T.K. Tay, R.H. Swendsen, Phys. Rev. Lett. {\bf
82}, 476 (1999); J.-S. Wang, R.H. Swendsen, J. Stat. Phys. {\bf
106}, 245 (2002)
\bibitem{berg-92} B.A. Berg, T. Neuhaus, Phys. Lett. B {\bf 276}, 249 (1991); B.A. Berg, T. Neuhaus, Phys. Rev. Lett. {\bf 68}, 9 (1992)
\bibitem{smith-95} G.R. Smith, A.D. Bruce, J. Phys. A {\bf 28}, 6623 (1995)
\bibitem{torrie-97} G.M. Torrie, J.-P. Valleau, J. Comput. Phys. {\bf 23}, 187 (1997)
\bibitem{swendsen-86} R.H. Swendsen, J.-S. Wang, Phys. Rev.
Lett. {\bf 57}, 2607 (1986)
\bibitem{geyer-91} C.J. Geyer, \textit{Computing Science and Statistics:
Proceedings of the 23rd Symposium on the interface}, ed. E.K.
Keramidas, Interface Foundation, Fairfax Station, New York, p. 156
(1991)
\bibitem{marinari-92} E. Marinari, G. Parisi, Europhys. Lett. {\bf 19}, 451 (1992)
\bibitem{lyubartsev-92} A.P. Lyubartsev, A.A. Martsinovskii, S.V. Shevkunov, P.N. Vorontsov-Velyaminov, J. Chem. Phys. {\bf 96}, 1776 (1992)
\bibitem{hukushima-96} K. Hukushima, K. Nemoto, J. Phys. Soc. Jpn. {\bf 65}, 1604 (1996)
\bibitem{marinari-98} E. Marinari, G. Parisi, J.J Ruiz-Lorenzo, \textit{Spin Glasses and Random Fields}, ed. A.P. Young,
Directions in Condensed Matter Physics, World Scientific,
Singapore, Vol. 12 (1998)
\bibitem{trebst-04} S. Trebst, D.A. Huse, M. Troyer, Phys. Rev. E {\bf 70}, 046701 (2004)
\bibitem{douarche-03} N. Douarche, F. Calvo, G.M. Pastor, P.J. Jensen, Eur. Phys. J. D {\bf 24}, 77 (2003)
\bibitem{troyer-03} M. Troyer, S. Wessel, F. Alet, Phys. Rev. Lett. {\bf 90}, 120201 (2003)
\bibitem{fytas-06} A. Malakis, N.G. Fytas, Phys. Rev. E {\bf 73}, 056114 (2006);
A. Malakis, N.G. Fytas, Phys. Rev. E {\bf 73}, 016109 (2006); N.G.
Fytas, A. Malakis, Eur. Phys. J. B {\bf 61}, 111 (2008)
\bibitem{schulz-05} B.J. Schulz, K. Binder, M. M\"{u}ller, Phys. Rev. E, {\bf 71} 046705 (2005)
\bibitem{reynal-05} S. Reynal, H.T. Diep, Phys. Rev. E {\bf 72}, 056710 (2005)
\bibitem{jayasri-05} D. Jayasri, V.S.S. Sastry, K.P.N. Murthy, Phys. Rev. E {\bf 72}, 036702 (2005)
\bibitem{trebst-05} S. Trebst, E. Gull, M. Troyer, J. Chem. Phys. {\bf 123}, 204501 (2005)
\bibitem{rathore-02} N. Rathore, J.J de Pablo, J. Chem. Phys. {\bf
116}, 7225 (2002); N. Rathore, T.A. Knotts, J.J. de Pablo, J.
Chem. Phys. {\bf 118}, 4285 (2003); N. Rathore, G. Yan, J.J de
Pablo, J. Chem. Phys. {\bf 120}, 5781 (2004); Q. Yan, R. Faller,
J.J. de Pablo, J. Chem. Phys. {\bf 116}, 8745 (2002)
\bibitem{shell-02} M.S. Shell, P.G. Debenedetti, A.Z. Panagiotopoulos, Phys. Rev. E {\bf 66}, 56703 (2002)
\bibitem{yamaguchi-01} C. Yamaguchi, Y. Okabe, J. Phys. A {\bf
34}, 8781 (2001); Y. Okabe, Y. Tomita, C. Yamaguchi, Comput. Phys.
Commun. {\bf 146}, 63 (2002)
\bibitem{carri-05} V. Varshney, G.A. Carri, Phys. Rev. Lett. {\bf 95}, 168304 (2005); G.A. Carri, R. Batman,
V. Varshney, T.E. Dirama, Polymer {\bf 46}, 3809 (2006)
\bibitem{calvo-03} F. Calvo, Mol. Phys. {\bf 100}, 3421 (2002); F. Calvo, P.
Parneix, J. Chem. Phys. {\bf 119}, 256 (2003)
\bibitem{tsai-07} S.-H Tsai, F. Wang, D.P. Landau, Phys. Rev. E {\bf 75}, 061108 (2007);
Braz. J. Phys. {\bf 38}, 6 (2008); D.T. Seaton, S.J. Mitchell,
D.P. Landau, Braz. J. Phys. {\bf 38}, 48 (2008); S.J. Mitchell,
F.C. Luiz Pereira, D.P. Landau, Braz. J. Phys. {\bf 38}, 1 (2008)
\bibitem{vorontsov-04} P.N. Vorontsov-Velyaminov, N.A. Volkov, A. Yurchenko,
J. Phys. A {\bf 37}, 1573 (2004); N.A. Volkov, P.N.
Vorontsov-Velyaminov, A.P. Lyubartsev, Phys. Rev. E {\bf 75},
016705 (2007)
\bibitem{poulain-06} P. Poulain, F. Calvo, R. Antoine, M. Broyer, P. Dugourd, Phys. Rev. E {\bf 73},
056704 (2006)
\bibitem{schulz-03} B.J. Schulz, K. Binder, M. M\"{u}ller, D.P. Landau, Phys. Rev. E {\bf
67}, 067102 (2003)
\bibitem{dayal-04} P. Dayal, S. Trebst, S. Wessel, D. W\"{u}rtz, M. Troyer, S. Sabhapandit, S.N. Coppersmith, Phys. Rev. Lett. {\bf 92}, 097201 (2004)
\bibitem{fytas-08} N.G. Fytas, A. Malakis, K. Eftaxias, J. Stat. Mech.: Theory Exp. (2008)
P03015;  A. Malakis, A.N Berker, I.A. Hadjiagapiou, N.G. Fytas,
Phys. Rev. E {\bf 79}, 011125 (2009); N.G. Fytas, A. Malakis,
Phys. Rev. E {\bf 81}, 041109 (2010)
\bibitem{alder-04} S. Alder, S. Trebst, A.K. Hartmann, M. Troyer, J. Stat.
Mech.: Theory Exp. (2004) P07008
\bibitem{zhou-05} C. Zhou, R.N. Bhatt, Phys. Rev. E {\bf 72}, 025701(R) (2005);
C. Zhou, T.C. Schulthess, S. Torbr\"{u}gge, D.P. Landau, Phys.
Rev. Lett. {\bf 96}, 120201 (2006)
\bibitem{parsons-06} D.F. Parsons, D.R.M. Williams, Phys. Rev. E {\bf 74}, 041804 (2006); D.F. Parsons, D.R.M. Williams, J. Chem. Phys. {\bf 124}, 221103 (2006)
\bibitem{netto-06} A.G. Cunha Netto, C.J. Silva, A.A. Caparica, R. Dickman, Braz. J. Phys. {\bf 36}, 619 (2006)
\bibitem{antypov-08} D. Antypov, J.A. Elliott, Macromolecules {\bf 41}, 7243 (2008)
\bibitem{wust-08} T. W\"{u}st, D.P. Landau, Comp. Phys. Commun. {\bf 179}, 124 (2008)
\bibitem{seaton-09} D.T. Seaton, T. W\"{u}st, D.P. Landau, Comp. Phys. Commun. {\bf 180}, 587
(2009); D.T. Seaton, T. W\"{u}st, D.P. Landau, Phys. Rev. E {\bf
81}, 011802 (2010).
\bibitem{taylor-09} M.P. Taylor, W. Paul, K. Binder, Phys. Rev. E {\bf 79}, 050801(R) (2009); M.P. Taylor, W. Paul, K. Binder, J. Chem. Phys. {\bf 131}, 114907 (2009)
\bibitem{fytas11} N.G. Fytas, P.E. Theodorakis, Phys. Rev. E {\bf 82}, 06201
(2010); P.E. Theodorakis, N.G. Fytas, Eur. Phys. J. B {\bf 81},
245 (2011)
\bibitem{wang-11} Z. Wang, X. He, J. Chem. Phys. {\bf 135}, 094902 (2011)
\bibitem{aizenman-89} M. Aizenman, J. Wehr, Phys. Rev. Lett. {\bf 62}, 2503 (1989); M. Aizenman, J. Wehr, Phys. Rev. Lett. {\bf 64}, 1311(E) (1990)
\bibitem{hui-89} K. Hui, A.N. Berker, Phys. Rev. Lett. {\bf 62}, 2507 (1989); K. Hui, A.N. Berker, Phys. Rev. Lett. {\bf 63}, 2433(E) (1989)
\bibitem{berker-93} A.N. Berker, Physica A {\bf 194}, 72 (1993)
\bibitem{chen-92} S. Chen, A.M. Ferrenberg, D.P. Landau, Phys. Rev. Lett. {\bf 69}, 1213 (1992); S. Chen, A.M. Ferrenberg, D.P. Landau, Phys. Rev. E {\bf 52}, 1377 (1995)
\bibitem{cardy-96} J. Cardy, J. Phys. A {\bf 29}, 1897 (1996)
\bibitem{cardy-97} J. Cardy, J.L. Jacobsen, Phys. Rev. Lett. {\bf 79}, 4063 (1997)
\bibitem{chatelain-98} C. Chatelain, B. Berche, Phys. Rev. Lett. {\bf 80}, 1670 (1998)
\bibitem{paredes-99} R. Paredes V., J. Valbuena, Phys. Rev. E {\bf 59}, 6275 (1999)
\bibitem{chatelain-01} C. Chatelain, B. Berche, W. Janke, P.E. Berche, Phys. Rev. E {\bf 64}, 036120 (2001)
\bibitem{fernandez-08} L.A. Fern\'{a}ndez, A. Gordillo-Guerrero, V. Mart\'{i}n-Mayor, J.J. Ruiz-Lorenzo, Phys. Rev. Lett. {\bf 100}, 057201 (2008)
\bibitem{harris-74} A.B. Harris, J. Phys. C {\bf 7}, 1671 (1974)
\bibitem{chayes-86} J.T. Chayes, L. Chayes, D.S. Fisher, T. Spencer, Phys. Rev. Lett. {\bf 57}, 2999 (1986)
\bibitem{dotsenko-95} V. Dotsenko, M. Picco, P. Pujol, Nucl. Phys. B {\bf 455}, 701 (1995)
\bibitem{jacobsen-98} J.L. Jacobsen, J. Cardy, Nucl. Phys. B {\bf 515}, 701 (1998)
\bibitem{MK-99} G. Mazzeo, R. K\"{u}hn, Phys. Rev. E {\bf 60}, 3823 (1999)
\bibitem{kenna} R. Kenna, D.A. Johnston, W. Janke, Phys. Rev. Lett. {\bf 96}, 115701 (2006); R. Kenna, D.A. Johnston,
W. Janke, Phys. Rev. Lett. {\bf 97}, 155702
(2006)
\bibitem{gordillo-09} A. Gordillo-Guerrero, R. Kenna, J.J. Ruiz Lorenzo, AIP Conf. Proc. {\bf 1198}, 42 (2009)
\bibitem{wang-90} J.-S. Wang, W. Selke, Vl.S. Dotsenko, V.B. Andreichenko, Physica A {\bf 164}, 221 (1990)
\bibitem{KP-94} J.-K. Kim, A. Patrascioiu, Phys. Rev. Lett. {\bf 72}, 2785 (1994); J.-K. Kim, A. Patrascioiu, Phys. Rev. B {\bf
49}, 15764 (1994); W. Selke, Phys. Rev. Lett. {\bf 73}, 3487
(1994); K. Ziegler, Phys. Rev. Lett. {\bf 73}, 3488 (1994); J.-K.
Kim, A. Patrascioiu, Phys. Rev. Lett. {\bf 73}, 3489 (1994); J.-K.
Kim, Phys. Rev. B {\bf 61}, 1246 (2000)
\bibitem{kuhn-94} R. K\"uhn, Phys. Rev. Lett. {\bf 73}, 2268 (1994)
\bibitem{suzuki-74} M. Suzuki, Prog. Theor. Phys. {\bf 51}, 1992 (1974)
\bibitem{gunton-75} J.D. Gunton, T. Niemeijer, Phys. Rev. B {\bf 11}, 567 (1975)
\bibitem{fisch-78} R. Fisch, J. Stat. Phys. {\bf 18}, 111 (1978)
\bibitem{malakis09} A. Malakis, A.N. Berker, I.A. Hadjiagapiou, N.G. Fytas, Phys. Rev. E {\bf 79}, 011125
(2009); A. Malakis, A.N. Berker, I.A. Hadjiagapiou, N.G. Fytas, T.
Papakonstantinou, Phys. Rev. E {\bf 81}, 041113 (2010)
\bibitem{guida-98} R. Guida, J. Zinn-Justin, J. Phys. A {\bf 31}, 8103 (1998)
\bibitem{peli} A. Pelissetto, E. Vicari, Phys. Rep. {\bf 368}, 549 (2002)
\bibitem{landau72} D.P. Landau, Phys. Rev. Lett. {\bf 28}, 449
(1972); A.N. Berker, M. Wortis, Phys. Rev. B {\bf 14}, 4946
(1976); M. Kaufman, R.B. Griffiths, J.M. Yeomans, M. Fisher, Phys.
Rev. B {\bf 23}, 3448 (1981); W. Selke, J. Yeomans, J. Phys. A
{\bf 16}, 2789 (1983); D.P. Landau, R.H. Swendsen, Phys. Rev. B
{\bf 33}, 7700 (1986); J.C. Xavier, F.C. Alcaraz, D. Pena Lara,
J.A. Plascak, Phys. Rev. B {\bf 57}, 11575 (1998)
\bibitem{stephen73} M.J. Stephen, J.L. McColey, Phys. Rev.
Lett. {\bf 44}, 89 (1973); T.S. Chang, G.F. Tuthill, H.E. Stanley,
Phys. Rev. B {\bf 9}, 4482 (1974); G.F. Tuthill, J.F. Nicoll, H.E.
Stanley, Phys. Rev. B {\bf 11}, 4579 (1975); F.J. Wegner, Phys.
Lett. {\bf 54A}, 1 (1975)
\bibitem{fox73} P.F. Fox, A.J. Guttmann, J. Phys. C {\bf 6},
913 (1973); T.W. Burkhardt, R.H. Swendsen, Phys. Rev. B {\bf 13},
3071 (1976); W.J. Camp, J.P. Van Dyke, Phys. Rev. B {\bf 11}, 2579
(1975); D.M. Saul, M. Wortis, D. Stauffer, Phys. Rev. B {\bf 9},
4964 (1974)
\bibitem{nightingale82} P. Nightingale, J. Appl. Phys. {\bf 53}, 7927 (1982)
\bibitem{beale86} P.D. Beale, Phys. Rev. B {\bf 33}, 1717 (1986)
\bibitem{jain80} A.K. Jain, D.P. Landau, Phys. Rev. B {\bf 22}, 445 (1980)
\bibitem{landau81} D.P. Landau, R.H. Swendsen, Phys. Rev. Lett. {\bf 46}, 1437 (1981)
\bibitem{care93} C.M. Care, J. Phys. A {\bf 26}, 1481 (1993)
\bibitem{deserno97} M. Deserno, Phys. Rev. E {\bf 56}, 5204 (1997)
\bibitem{blote03} Y. Deng, H.W.J. Bl\"{o}te, Phys. Rev. E {\bf 68}, 036125 (2003)
\bibitem{silva06} C.J. Silva, A.A. Caparica, J.A. Plascak, Phys. Rev. E {\bf 73}, 036702 (2006)
\bibitem{kutlu06} A. \"{O}zkan, N. Sefero\u{g}lu, B. Kutlu, Physica A {\bf 362}, 327 (2005)
\bibitem{hasen10} M. Hasenbusch, Phys. Rev. B {\bf 82}, 174433 (2010)
\bibitem{grollau01} S. Grollau, E. Kierlik, M.L. Rosinberg, G. Tarjus, Phys. Rev. E {\bf 63}, 041111 (2001)
\end{thebibliography}
\end{document}